\documentclass[3p,times]{elsarticle}

\usepackage{ecrc}

\volume{00}

\firstpage{1}

\journalname{Procedia Computer Science}

\runauth{Z. Jin et al.}

\jid{procs}

\jnltitlelogo{Procedia Computer Science}

\usepackage{amssymb}

\begin{document}

\begin{frontmatter}



\dochead{International Conference on Computational Science, ICCS 2010}

\title{High-performance astrophysical visualization using Splotch}


\author{Zhefan Jin}
\ead{zhefan.jin@gmail.com}
\address{School of Creative Technologies, University of Portsmouth, Winston Churchill Avenue, Portsmouth, United Kingdom}

\author{Mel Krokos}
\ead{mel.krokos@port.ac.uk}
\address{School of Creative Technologies, University of Portsmouth, Winston Churchill Avenue, Portsmouth, United Kingdom}

\author{Marzia Rivi}
\ead{m.rivi@cineca.it}
\address{CINECA, Via Magnanelli 6/3, Casalecchio di Reno, Italy}

\author{Claudio Gheller}
\ead{c.gheller@cineca.it}
\address{CINECA, Via Magnanelli 6/3, Casalecchio di Reno, Italy}

\author{Klaus Dolag}
\ead{kdolag@mpa-garching.mpg.de}
\address{Max-Planck-Institut f\"ur Astrophysik, Karl-Schwarzschild Strasse 1, Garching bei M\"unchen, Germany}

\author{Martin Reinecke}
\ead{martin@mpa-garching.mpg.de}
\address{Max-Planck-Institut f\"ur Astrophysik, Karl-Schwarzschild Strasse 1, Garching bei M\"unchen, Germany}

\begin{abstract}

The scientific community is presently witnessing an unprecedented growth in 
the quality and quantity of data sets coming from simulations and real-world experiments. 
To access effectively and extract the scientific content of such large-scale data sets 
(often sizes are measured in hundreds or even millions of Gigabytes) appropriate tools are needed. 
Visual data exploration and discovery is a robust approach for rapidly and 
intuitively inspecting large-scale data sets, e.g. for identifying new features and 
patterns or isolating small regions of interest within which to apply 
time-consuming algorithms. This paper presents a high performance parallelized implementation of 
Splotch, our previously developed visual data exploration and discovery algorithm for 
large-scale astrophysical data sets coming from particle-based simulations. 
Splotch has been improved in order to exploit modern massively parallel 
architectures, e.g. multicore CPUs and CUDA-enabled GPUs. We present performance 
and scalability benchmarks on a number of test cases, demonstrating the ability 
of our high performance parallelized Splotch to handle efficiently large-scale data sets, 
such as the outputs of the Millennium II simulation, the largest cosmological simulation ever performed.

\end{abstract}


\end{frontmatter}


\section{Introduction}
\label{intro}

Nowadays the technological advances in instrumentation and computing 
capability impact profoundly on the dramatic growth in the quality and 
quantity of astrophysical data sets obtained from observational instruments,
e.g.\ sky surveys \cite{sdss}, \cite{lofar}, or large-scale numerical
simulations, e.g.\ the Millennium II simulation \cite{2009MNRAS.398.1150B}.

The main characteristic of modern astrophysical data sets is extremely large
sizes (in the order of hundreds of Gigabytes) requiring storage in extremely
large-scale distributed databases. The forthcoming next-generation astrophysical 
data sets are expected to exhibit massively large sizes (in the order of hundreds
of Terabytes), e.g.\ \cite{lsst}.
To obtain a comprehensive insight into modern astrophysical data sets, astronomers
employ sophisticated data mining algorithms, often at prohibitively high computational costs. 
Visual data exploration and discovery tools are then exploited in order to rapidly 
and intuitively inspect very large-scale data sets to identify regions of interest
within which to apply time-consuming algorithms. Such tools are based on a 
combination of meaningful data {\it visualizations} and user interactions with them.  

This apporoach
can be a very intuitive and ready way of discovering and 
understanding rapidly new correlations, similarities and data patterns. 
For on-going processes, e.g. a numerical simulation in progress,  
visual data exploration and discovery allow constant monitoring and - if 
anomalies are discovered - prompt correction of the run, thus saving valuable 
time and resources. 

The data exploration tools traditionally employed by astronomers 
are limited either 
to processing and displaying of 2D images (see, e.g., \cite{iraf},
\cite{midas}, \cite{sao},
\cite{gaia}) or to
generation of meaningful 2D and 3D plots (e.g.\ \cite{gnuplot},
\cite{supermongo}, \cite{idl}).
 

To overcome the shortcomings of traditional tools, a new generation of software 
packages is now emerging, providing astronomers with robust instruments in the context 
of large-scale astrophysical data sets (e.g.\ \cite{paraview}, \cite{aladin},
\cite{topcat}, \cite{visivo1} and \cite{visivo2}, 
\cite{3dslicer}, \cite{splash} and \cite{visit}). 
The underlying principles are exploitation of high performance architectures 
(i.e.\ multicore CPUs and powerful graphics boards), interoperability
(different applications can operate simultaneously on
shared data sets) and collaborative workflows (permitting several users to
work simultaneously for exchanging information and visualization experiences).

%

This paper describes a high performance implementation of 
Splotch \citep{2008NJPh...10l5006D}, our previously developed ray-tracing 
algorithm for effective visualization of large-scale astrophysical data sets coming 
from particle-based computer simulations. N-Body simulations constitute 
prime examples of particle-based simulations, typically associated with very 
large-scale data sets, e.g.\ the Millennium II simulation \citep{2009MNRAS.398.1150B}.
This is a simulation of the evolution of a meaningful fraction of the universe 
by means of 10 billion fluid elements ({\it particles}) interacting with each other 
through gravitational forces. The typical size of a snapshot of the Millennium II 
simulation is about 400 Gigabytes representing a particle's ID, position and velocity
together with additional properties, e.g.\ local smoothing length, density and 
velocity dispersion. For further details on the Millennium II simulation and
other works about the visualization of its data sets, the reader
is referred to \citep{2009MNRAS.398.1150B}, \cite{fraedrich2009} and \cite{Szalay2008}.

The fundamentals and the traditional sequential operation of Splotch 
are reviewed in section 2. Section 3 discusses 
our strategy for parallelizing Splotch based on different approaches that 
are suitable for a variety of underlying architecture configurations. 
Our implementations are Single Instruction Multiple Data (SIMD) designs 
founded on the MPI library \cite{mpi} in order to support distributed multicore 
CPUs and 
CUDA \cite{cuda} for exploiting not only currently available but also forthcoming 
next-generation multiple GPUs. The advantage of adopting several parallelization 
solutions is that we can deploy them simultaneously on hybrid architectures, 
e.g.\ mixed hardware architectures consisting of a large number of multicore CPUs 
and CUDA-enabled GPUs. Benchmarks for our parallelization designs and a discussion 
on the Millenium II visualization are presented in section 4. Finally section 5 
outlines a summary of our work and includes pointers to future developments.

\section{The Splotch Algorithm}
\label{splotch}

\begin{figure}
\begin{center}
\includegraphics[width=0.40\textwidth]{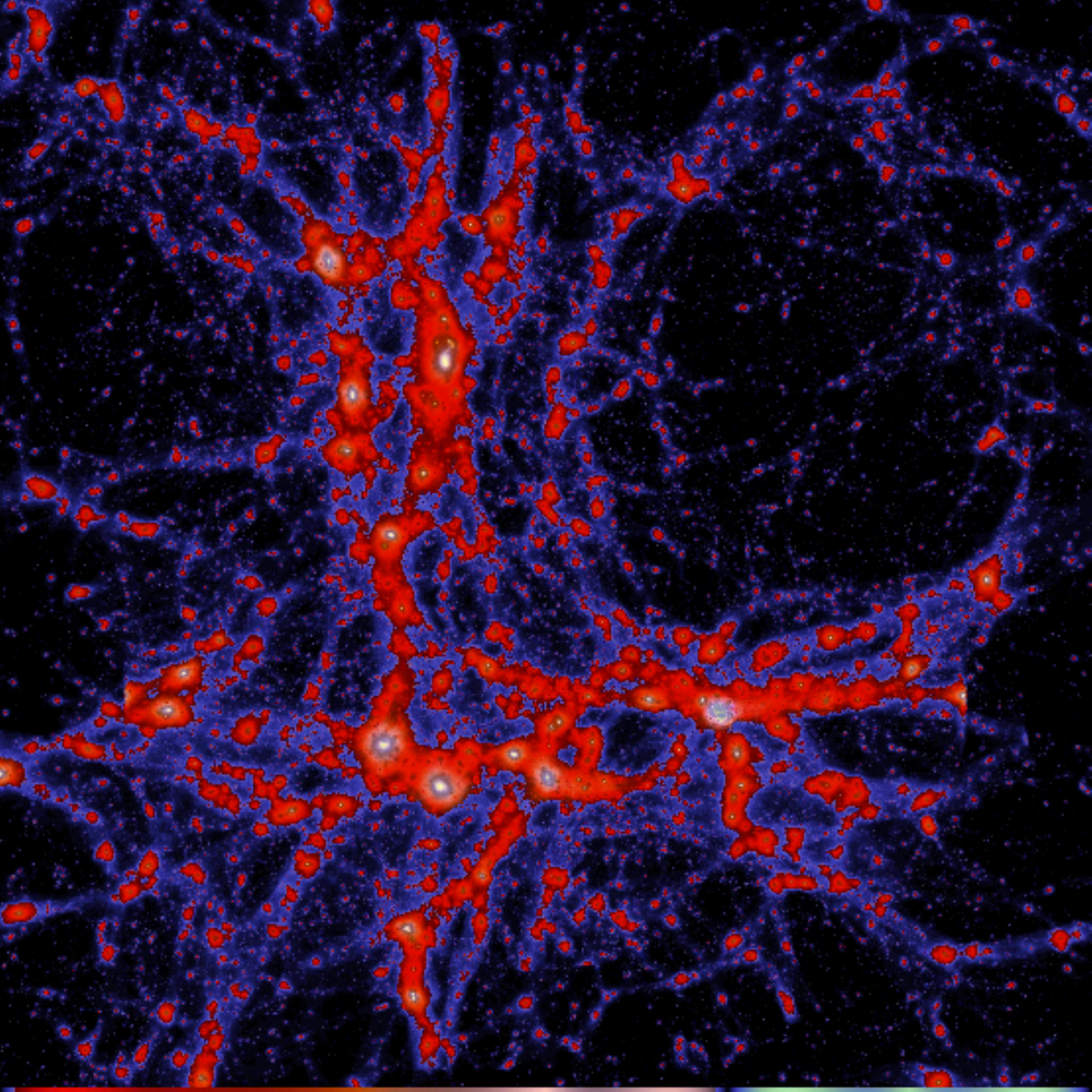}
\includegraphics[width=0.40\textwidth]{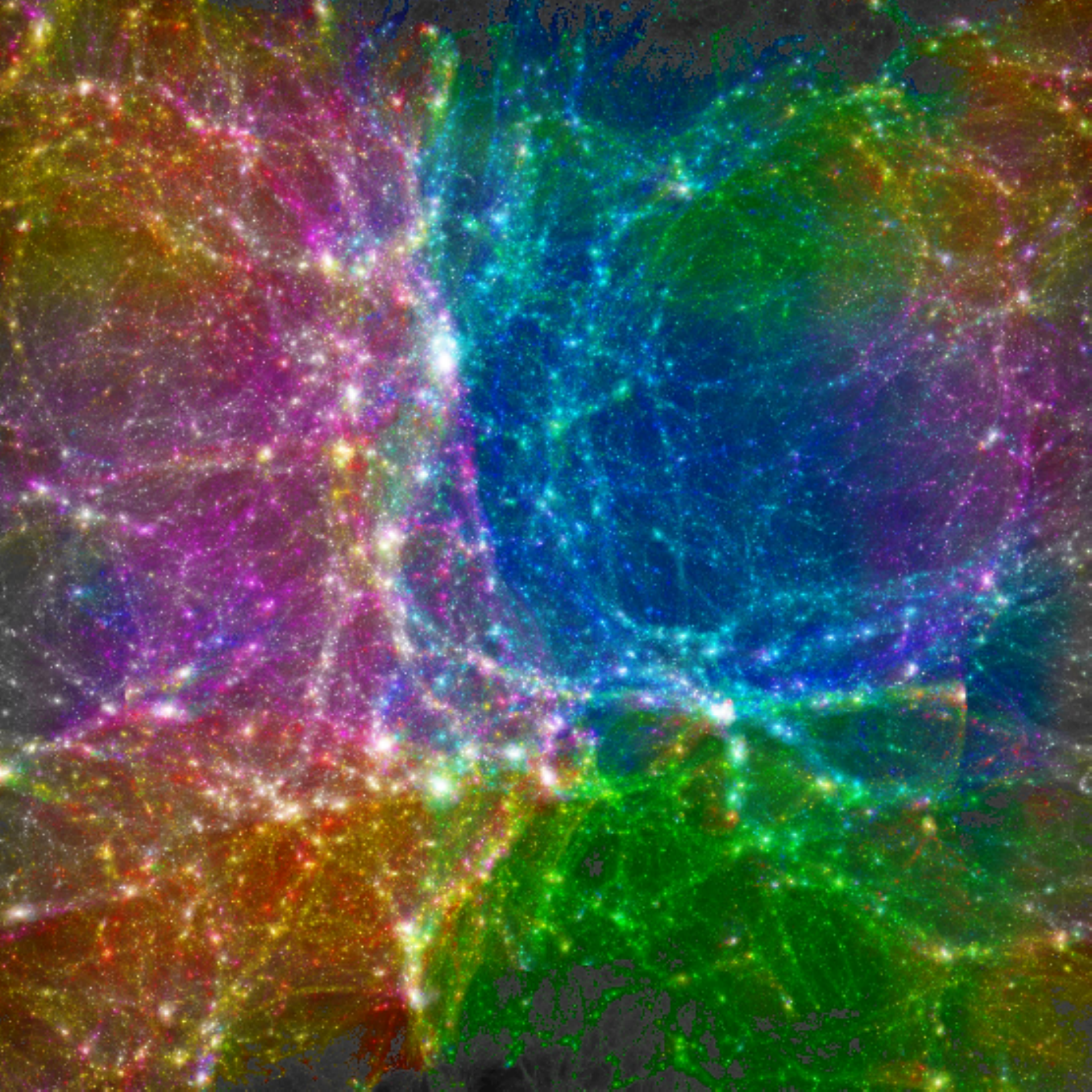}
\end{center}
\caption{A visualization  of the Millennium II simulation \citep{2009MNRAS.398.1150B}. 
The color transfer function uses a particle's velocity dispersion (left) and 3D velocity (right).}\label{mil2}
\end{figure}

The rendering algorithm of Splotch is designed to handle
point-like particle distributions. Such tracer particles can be smoothed
to obtain a continuous field, which is rendered based
on the following assumptions:

\begin{itemize}
\item
The contribution to the matter density by every particle can
be described by a Gaussian distribution 
$\rho_p(\vec r)=\rho_{0,p}\exp(-r^2/\sigma_p^2)$.
In practice, it is much more handy to have a compact support of the
distribution, and therefore the distribution is set to zero at a given
distance $f\cdot\sigma_p$, where $f$ is a proper multiplicative factor. 
Therefore rays passing
the particle at a distance larger than $f\cdot\sigma_p$ will be
unaffected by the particle's density distribution.

\item 
We use three ``frequencies'' to describe the red, green and blue
components of the radiation, respectively. These are treated independently.

\item
The radiation intensity $\bf{I}$ (treated 
as a vector with r,g and b components) along a ray through the simulation
volume is modeled by the well known radiative transfer equation
\begin{equation}
\frac{d\bf{I}(x)}{dx}=(\bf{E}_p-\bf{A}_p\bf{I}(x))\rho_p(x),
\end{equation}
which can be found in standard textbooks \cite{1991par..book.....S}.
Here, $\bf{E}_p$ and $\bf{A}_p$ describe the strength of radiation emission and absorption
for a given particle for the three rgb-colour components. In general it is recommended to
set $\bf{E}_p=\bf{A}_p$, which typically produces visually appealing images. This is presently a 
necessary setting for Splotch, in order to reduce the complexity of some aspects of its parallel
implementation. This constraint will be eliminated in the next releases of the code.
If a scalar quantity is chosen (e.g.\ the particle temperature,
density, velocity dispersion, etc.), the mapping to the three components of $\bf{E}$ and $\bf{A}$ (for red, green and blue)
is typically achieved via a transfer function, realized by a colour look-up table or palette, which can
be provided to the ray-tracer as an external file to allow a maximum of flexibility. If a
vector quantity is chosen (e.g.\ velocity, magnetic field, etc.), the three components of the vectors
can be mapped to the three components of $\bf{E}$ and $\bf{A}$ (for red, green and blue). In addition 
to the color, the optical depth of each particle can be also modulated proportionally to another
scalar property (e.g.\ density, etc.).
\end{itemize}

%

Further details on the Splotch rendering algorithm can be found in \citep{2008NJPh...10l5006D}. 
Figure \ref{mil2} shows a visualization example of a large simulation
containing 10 billion particles.

\section{Parallel Implementation}
\label{parallel}

\begin{figure}
\begin{center}
\includegraphics[width=1.00\textwidth]{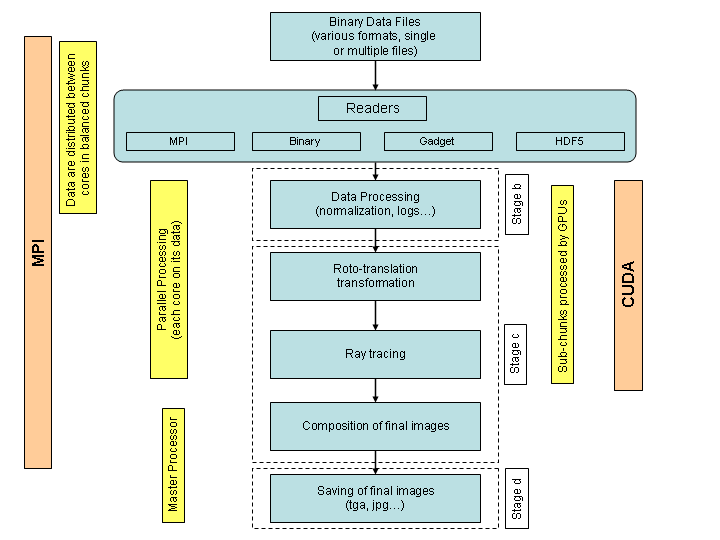}
\end{center}
\caption{Workflow of the parallelized Splotch algorithm.}\label{flow}
\end{figure}

The Splotch algorithm operational scenario consists of a number of 
stages summarized as follows: a) read data from one or more files; b) process data (e.g.\ for normalization); 
c) render data and d) save the final image. All these steps can be parallelized using a
Single Instruction Multiple Data (SIMD) approach. This involves distributing the data 
in a balanced way between different computing elements (or processing units) 
and each computing element performing the same operations on its associated subset of data. 
Our parallelization has been achieved using different approaches suitable for a variety of 
underlying hardware architectures and software environments. Figure \ref{flow} outlines the 
overall workflow of the parallelised Splotch algorithm.

The MPI library has been used to define the overall data and work 
distribution. Data are read in chunks of the same (or similar) size by each processor
in the MPI pool. Then the same work is performed for steps (b) and (c) on each data chunk by the
corresponding processor. The final image is generated and saved only by the root processor. 
This is, in fact, a light task, which does not require any kind of parallel implementation.
The work accomplished in steps (b) and (c) can be further split, exploiting 
multicore shared memory processors or modern graphics boards, by means of
an OpenMP based approach or exploitation of the CUDA programming 
environment. 

These different parallel approaches can be used separately, 
if the available computing system fits only one of the available configurations, or jointly. 
For instance, on a single core PC with an NVIDIA graphics card, only the CUDA based
parallelization strategy can be activated and exploited. On a multicore RVN  
node both MPI and CUDA can be used. This makes our parallel Splotch code 
extremely flexible, portable, efficient and scalable. 
More details of our parallel implementations are presented in the rest of this section. 
Test results are discussed in Section 4.

\subsection{MPI Implementation}
\label{mpi}

Once the data is distributed among the processors, all the remaining operations 
are performed locally and further communication is not needed, until the generation
of the final display. Each MPI process uses the assigned 
data to produce its own partial image. At the end, all the partial contributions are merged by means of a
collective reduction operation producing the final image. As the data load stage is the crucial step for balancing the overall workload and fast reading
data from the disk, we paid specific attention to efficient implementation of this functionality.

The adoption of MPI I/O based functions represents the ideal solution for obtaining
a high-performance, scalable data input utility.
With this approach each process has a different view of individual files. 
that allows simultaneous and collective
writing/reading of non-contiguous interleaved data. 
Our implementation assumes that data are organized in the input file according to a block structure, 
where each block contains a single information for 
all $N$ particles. Therefore we have as many contiguous blocks
as the number $n$ of properties given for each particle, and we can see them as a 2-dimensional array $A$ 
of $n \times N$ float elements. 
Then, we have defined the MPI I/O filetype as a simple 2-dimensional subarray of $A$ 
of size $n \times N/nprocs$.

To support high performance computing environments where MPI I/O is not available, 
we have also provided two standard MPI binary readers based on standard {\tt fstream} functions.
Data to be read are equally distributed among processes and each one reads simultaneously
their own portion of data by a direct access operation. 
%
In all readers an endianness conversion is also performed if required.

\subsection{CUDA Implementation}
\label{cuda}
Nowadays Graphics Processing Units (GPUs) can offer a means of increased performance 
(often substantially) in the context of computationally intensive scientific applications by exploiting high speed underlying ALUs and stream data-parallel organization. 
The Compute Unified Device Architecture (CUDA) introduced by NVIDIA offers access to highly parallellized modern GPU architectures via a 
simplified C language interface. 
A thread on the GPU is extremely lightweight compared to CPU threads, so changing context 
among threads is not a costly operation. The minimum 
data chunks processed by a CUDA {\it multiprocessor} are numbers of threads ({\it warps}) handled in groups 
as {\it blocks} and {\it grids} so that GPU-executed functions can exploit large numbers of threads. 
CUDA executes blocks sequentially
in case of limited hardware resources, but reverts to parallel execution for large numbers 
of processing units.  The resulting code can thus target simultaneously entry-level, 
high-end or even next-generation GPUs. Further details on CUDA can be found on \cite{cudaprogguide}.


As soon as data is loaded in the memory of each processing unit (defined as a multicore CPU together with a bundle of associated GPUs), our CUDA approach can be combined with the MPI parallelization strategy outlined in Section 3.1. At that point in fact, processing units can be regarded as completely independent of each other, and CUDA can be exploited by determining each parallel task based on a single particle, that is a single particle is {\it processed} and {\it displayed} by a single CUDA thread. 
The processing involves normalizing of some particle values. 
The displaying involves transformation into screen coordinates, assigning of colours and rendering for determining screen areas affected by individual particles and subsequently combining them for final imaging.

The granularity during processing, transformation and colorization is more or 
less fixed among different particles. However during rendering it 
can vary considerably depending upon the number of screen pixels influenced 
by individual particles. As a worst case scenario consider two particles influencing 
all screen pixels and a single screen pixel respectively. 
Assuming they are handled by the same warp (this is determined by CUDA automatically), 
they are then scheduled to execute simultaneously. This unbalanced granularity 
can compromise significantly overall execution times.


To alleviate this situation we follow a 'split particles' strategy 
dividing large computational tasks into smaller average ones.  For any particle 
influencing a number of pixels that is larger than a threshold value, the relevant 
computational task is sub-divided into multiple ones, each associated with a subset 
of the original number of pixels. The threshold value 
can be determined in advance and given as input to Splotch. To our experience a threshold close to 
the average of the width and height of the display window works satisfactorily. 
A shortcoming of this is the computational cost when doing the splitting; for an increased number of particles 
more threads are required. Execution of the splitting algorithm 
and memory copying among host and device involve additional costs. 
Nevertheless our results demonstrate improved timings (see Table 2).
The pseudo-code below summarizes the overall workflow of our CUDA paralellization approach.

\newpage
\begin{verbatim}
while  ( not all particles are rendered )
{
       find subset S(i) of particle array;
A:     call device to render S(i);
       if  ( S(i) is not first subset )
       { 
B:          combine with F(i-1), the output of S(i-1) in fragment buffer;
       }
C:     copy fragment buffer from device to host;
       if  ( S(i) is the last subset )
       {
             combine with F(i);
       }
       i++;
}
\end{verbatim}

The instruction A is the render computation executed on the graphics board 
in parallel with execution of instruction B, while 
the combination operation is carried out by the CPU. The results of A and B are merged by instruction C. 
Our test results indicate that overall times required for performing combination operations are mostly
contained within the times demanded by render computations.


\section{Benchmarks}

The parallelized Splotch code has been tested on low-end and high-end hardware 
architectures using several test data sets in order to investigate applicability of
different parallelization approaches.

Exploiting the high portability of our code we could perform our tests on a 
5000 cores UNIX-AIX  SP6 system (referred to as SP6) and a Windows XP PC (referred to as Win). 
The SP6 system is a cluster of 168 Power6 575 computing nodes with 32 cores
and a memory of 128 Gigabytes per node. The Win system is an Intel Xeon X5482 3.2 GHz CPU with 
two NVIDIA Quadro FX 5600 graphics boards. Our target was to test parallelized
versions of Splotch on computing systems of different sizes and target applications, 
from a standard PC, where small to medium data sets can be used, up to high
performance platforms for handling very large data sets.

We used several benchmark data sets for our testing. The first
few data sets are derived from a cosmological N-Body simulation of more than 850 million
particles characterized by spatial information together with velocities, mass density 
and smoothing length, for defining the size of the region influenced by the properties 
of each particle when deploying Splotch. We randomly extracted data sets containing
1, 10 and 100 million particles. These are indicated as 1M, 
10M and 100M tests. We also employed the data set 850M containing the entire simulation.
Such variety of data set sizes is required to match the memory available on
the different computing systems used for our testing.
Our most challenging data set comes from the Millennium II simulation \cite{2009MNRAS.398.1150B}
and consists of 10 billion particles. The data files employed are pure binaries organized in such a way that different
quantitites are stored consecutively (e.g.\ $x$ coordinates of all particles, 
then $y$ coordinates and so on) as expected by the MPII/O based reader.
 
We provide a rough estimate of the errors associated with our measures only for the Millennium Run 
(the most challenging). The benchmarks, in fact, were performed on production systems, where 
random fluctuations in the workload can affect the results. This is expecially true 
for I/O, which is shared between a number of users and jobs. 
Therefore, we chose to present the best performance obtained in each test, focusing on the 
scalability of the code, more than on the absolute result. Anyway, the sample 
error bars presented in Figure \ref{cpu_scaling},
show that the performance has negligible fluctuations, slightly larger, as 
expected, for the I/O operations.

\subsection{MPI Benchmarks}

\begin{figure}
\begin{center}
\includegraphics[width=0.45\textwidth]{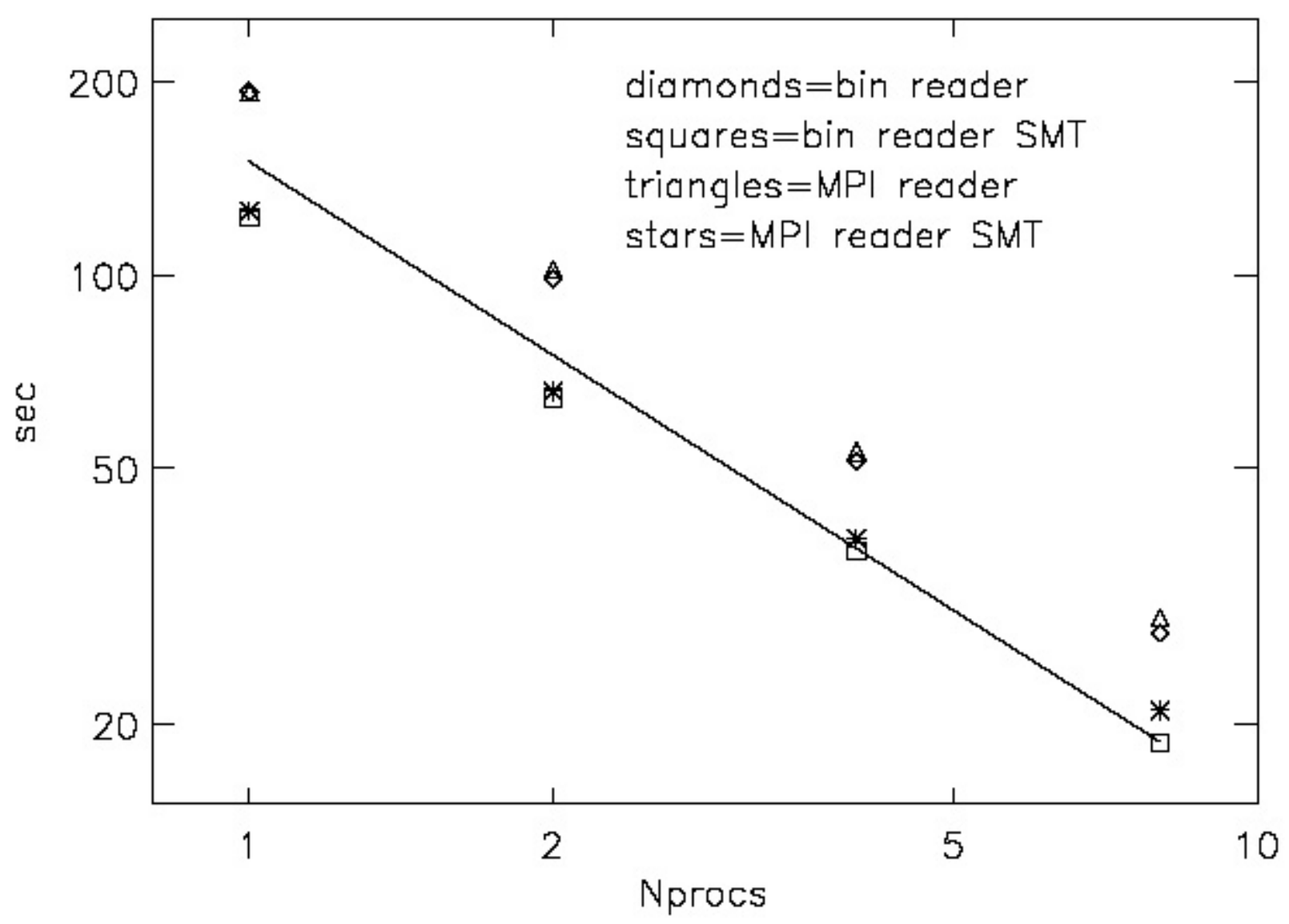}
\includegraphics[width=0.45\textwidth]{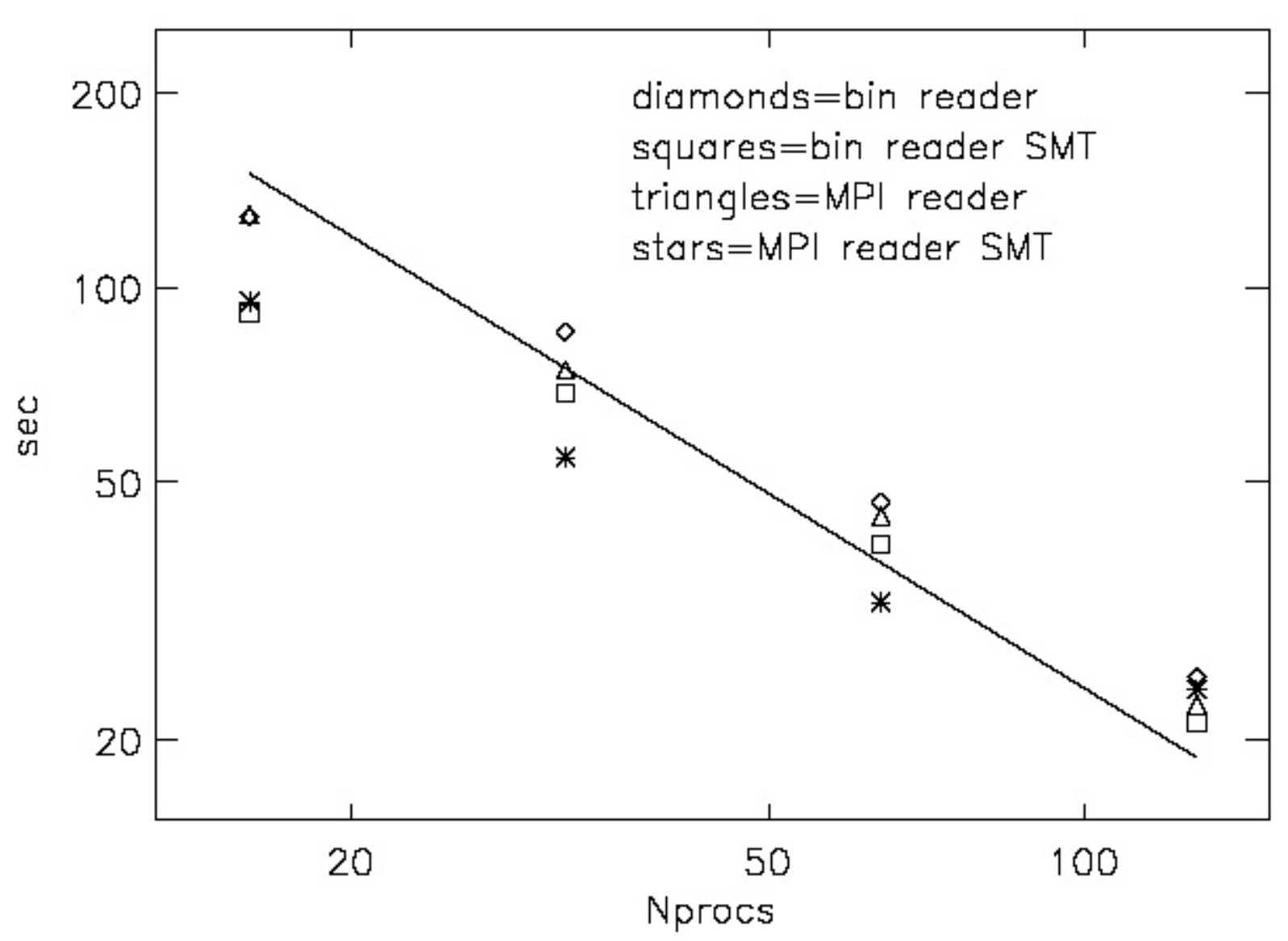}
\end{center}
\caption{Scaling of the CPU time (total wallclock time) with the number of MPI threads 
used for vizualizing 100 million (left panel)
and 850 million particles (right panel) for an 800x800 pixel display.
Results for binary and MPI readers, in ST and SMT modes are shown.}\label{mpi100M}
\end{figure}

Our test results on the SP6 platform for data sets 100M and 850M
are presented in Figure \ref{mpi100M}.
Power6 cores can schedule for execution two processes (or threads) in the same clock cycle, 
and it is also possible to use a single core as two virtual CPUs. This mode of Power6 
is called Simultaneous Multi-Threading (SMT), to distinguish it from 
the standard Single Thread mode (ST).
Deploying SMT notably improves the performance 
of processing and displaying in Splotch by reducing the execution time up to  
$30\%$.

The 100M test (Figure \ref{mpi100M}, left panel), allows us 
to run on one processor fitting its memory. 
This can be considered a proper estimate of the sequential performance of the code, since, on a single processor the 
MPI calls overhead is negligible. 
The maximum number of processors for this test is set to 8, since the usage of 
more processors is inefficient, due to the small size 
of the data chunks assigned to each processor and the increasing overhead of communication.
On a single processor we get a best performance of about 123 seconds, which means that 
we can load and process approximatively 1 particle per $10^{-6}$ seconds. 
In all cases the code scales almost linearly up to 8 processors, beginning to lose
efficiency between 4 and 8 processors due to the reasons previously explained. 

The larger test, 850M (Figure \ref{mpi100M}, right panel), 
allows us to perform a more extensive scalability test, 
exploring the range between 16 and 128 processors. This test confirms that parallel
Splotch can process about 1 particle per microsecond per process. For this data set the scalability
is demonstrated up to 128 processors in the ST mode. The SMT configuration,
while producing the best absolute performances, seems to have a more limited scalability. 
This requires more investigation, but our anticipation is that the most likely reason for this is processor 
architectural features rather than our code.

The pure binary and the MPI readers lead to similar performances. This is due to the features 
of our data set, which is characterized by one dimensional data arrays. This means
that the parallel reading functions can read large chunks of contiguous data 
in a single operation. However improvements due to MPI2 functions appear when 
processes read large chunks of data. So good performances are expected as size of data increases. 
Moreover when the number of processes is high, the MPI I/O reader performs and scales 
better than the other one. Further improvements should emerge when multidimensional 
arrays are considered and the access data pattern is more complex. 
In these cases, collective MPI I/O reading could provide an effective speed-up for data loading.

\subsection{CUDA Benchmarks}

\begin{table}
\caption{The performance timings (secs) obtained with our CUDA implementation of Splotch (indicated by yes) compared to timings obtained using the standard sequential implementation of Splotch (indicated by no) for benchmark data sets 1M and 10M.}
\begin{center}
\begin{tabular}{|l|l|l|l|l|}
\hline
	& setup/read & 	display & 	write & 	total \\
\hline
1M (no) & 	1.1259 & 	0.6982 & 	0.1860 & 	2.0103 \\
\hline
1M (yes) & 	1.0916 & 	0.8411 & 	0.1840 & 	2.1170 \\
\hline
10M (no) & 	10.7064 & 	6.1376 & 	0.1842 & 	17.0284 \\
\hline
10M (yes) & 	10.1411 & 	4.8261 & 	0.1865 & 	15.1537 \\
\hline
\end{tabular}
\end{center}
\end{table}

\begin{table}
\caption{The performance timings (secs) obtained with our CUDA implementation of Splotch. 
The number of particles in the data sets employed for our experimental results is
370,852 (small), 2,646,991 (medium) and 16,202,527 (large) respectively.}

\begin{center}
\begin{tabular}{ | l | l | l | l | l | p{3cm} |}
\hline  
  & \multicolumn{2}{|c|}{CPU}	& \multicolumn{3}{|c|}{GPU} \\
\hline
  data set & disp. t. & tot. t. & disp. t. & tot. t. & disp. t. no split \\
\hline  
  small &	12.6419	& 13.0144	& 7.3885	& 7.7668	& 12.4894 \\
\hline
  medium &	18.5443 &	19.7190 &	11.3855 &	12.7118 &	16.6905 \\
\hline
  large	& 31.8976 &	39.3234 &	19.7792 &	27.0321	& 24.6305\\
\hline
\end{tabular}
\end{center}
\end{table}

\begin{figure}
\begin{center}
\includegraphics[width=1.0\textwidth]{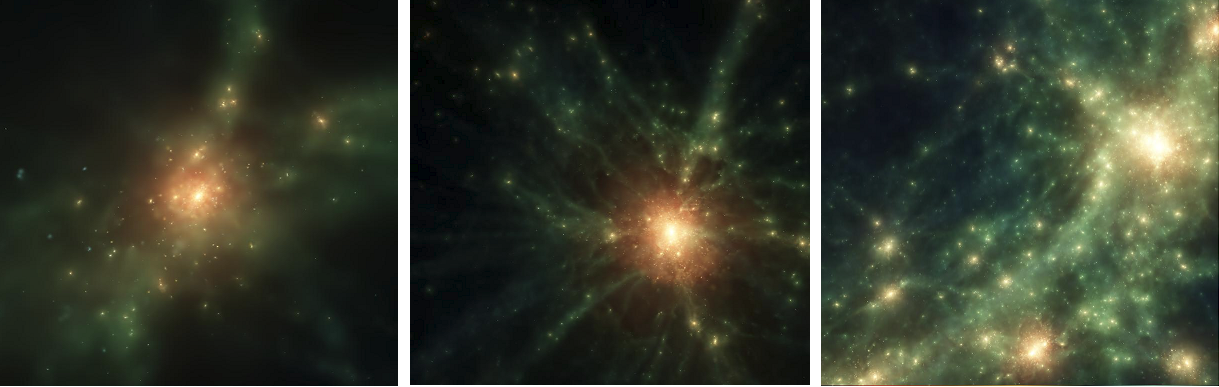}
\end{center}
\caption{Sample renderings of small (left), medium (middle) and large (right) data sets.}\label{cudafig}
\end{figure}

The performance timings for benchmark data sets 1M and 10M using our CUDA
implementation under the Win configuration described earlier are presented in Table 1. 
The overall performance for data set 1M is somewhat degraded when CUDA is employed.
The main reason is the fact that for data set sizes up to this order of magnitude the
costs associated with standard CUDA operations (e.g.\ initializing CUDA runtime 
or data copying from host to device and vice-versa) are non-negligible. 
As the size of our benchmark data set becomes an order of magnitude larger,
using CUDA results in 21.3\% performance gains (computed as
: Gain = (TimeWithoutCUDA - TimeWithCUDA) / TimeWithoutCUDA). 
Our anticipation is that further increasing the order of magnitude of a data set's 
size would result in further performance gains. Our initial experiences with several 
other data sets so far indicate that our CUDA implementation offers maximum gains
when there is a large display calculation involved, that is when individual particles 
influence relatively large areas on the screen. This is indeed the case for typical 
display scenarios (see Figure \ref{cudafig}). Table 2 shows that our CUDA parallelization improves the 
performance of Splotch. The 'split particles' strategy (see Section 3.2) further 
improves performance timings.
 

\subsection{Millennium II Visualization}
\label{mII}

The visualization of the outputs of the Millennium II simulation \cite{2009MNRAS.398.1150B} 
represents our most challenging benchmark as it requires the processing of 10 billion particles 
simultaneously. To do this at least 300 Gigabytes of memory are necessary, only available 
on HPC platforms such as the SP6 system.
In our performance results we have used the final output of the simulation, producing 
high resolution images of 3200x3200 pixels. A minimum number of 128 processors 
is necessary to process the entire data set.

For this benchmark, we could not use our MPI reader, due to a serious limitation related 
to the MPI I/O API (usage of 32 bits counters which do not allow us to handle the large-scale data sets of the Millennium).
We will work to overcome this problem in the future versions of Splotch. 
At the moment, we used instead the built-in parallel reader derived by the original Gadget 2 code 
(used for the simulation - \cite{gadget}), 
which allows to access directly data files as they where saved by the simulation code. 

\begin{figure}
\begin{center}
\includegraphics[width=0.49\textwidth]{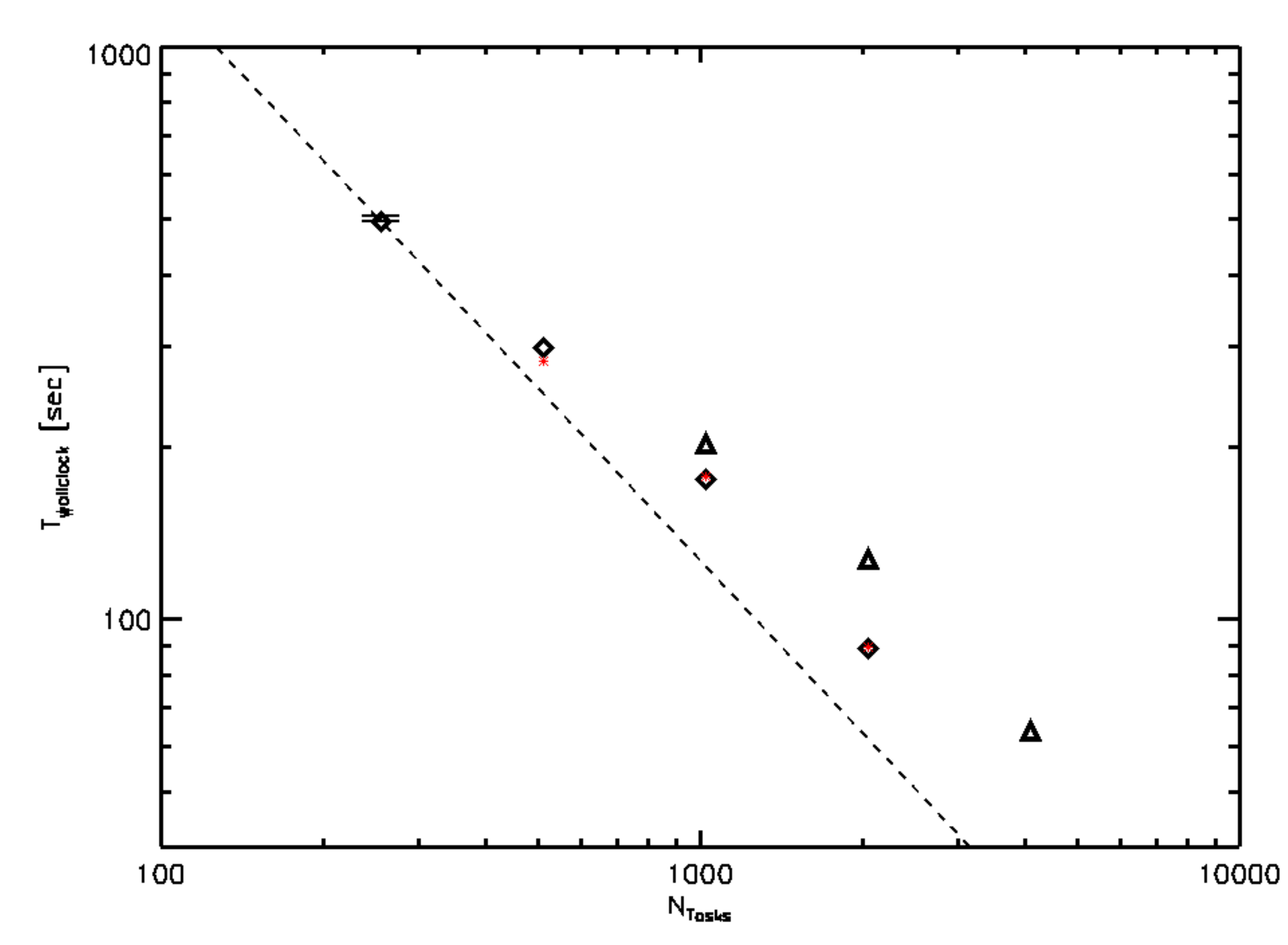}
\includegraphics[width=0.49\textwidth]{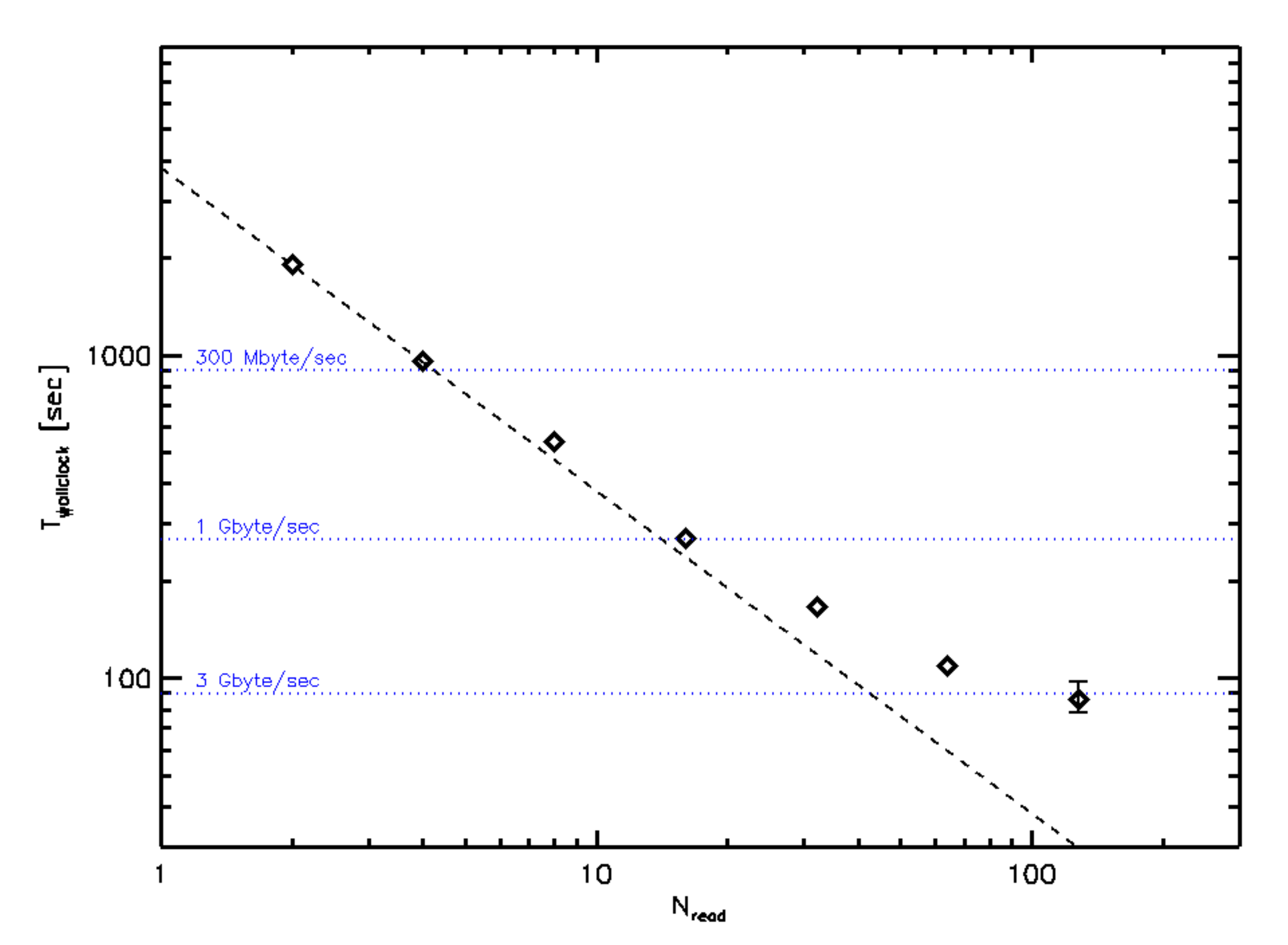}
\end{center}
\caption{Scaling of the CPU time with the number of MPI threads used 
for the Millennium II simulation data. 
The left panel shows the total wallclock time minus the time needed for 
reading, the right panel the read-data time. 
The dashed line indicates the expectation for an ideal scaling.  
The diamonds represent runs in ST mode,
the triangles indicate the SMT configuration.
The 128 processors point shows also the error bar, calculated as the variance of 13 repeated realizations. 
The processing time error is extremely small and the error bar is within the data point.
}\label{cpu_scaling}
\end{figure}

Examples of the Millennium II simulation rendered images are shown in Figure \ref{mil2}.
In Figure \ref{cpu_scaling}, we present the results of the benchmarks, in ST and SMT modes.
The left panel shows the data processing time from 128 to 2048 MPI processes in ST mode and from
256 to 4096 threads for SMT. The right panel shows instead the data read time, using 
the Gadget 2 reader. In this case, the read process is performed by a subset of processors,
belonging to a dedicated MPI sub-communicator, which then scatters the loaded data to the complete
processors pool. The tests confirm the scalability properties of the code even when a large number
of threads is used. The reader's curve tends to flatten toward 100 processors, achieving 
the physical bandwidth limit of the underlying GPFS based storage system. 

\section{Summary and Future Developments}
\label{conclusions}


In this paper we have described a high performance parallel implementation of Splotch 
able to execute on a variety of high performance computing architectures. This is due 
to its hybrid nature exploiting multi-processor systems adopting an MPI based approach,
multi-core shared memory processors exploiting OpenMP, and modern CUDA enabled graphics 
boards. This allows to achieve extremely high performance overcoming the typical memory 
barriers posed by small personal computing systems, commonly adopted for visualization. 
Finally, as parallel Splotch is implemented in ISO C++ and is completely self-contained
(in other words it does not require any complementary library apart from MPI, OpenMP and CUDA), 
the code is highly portable and compilable over a large number of different 
architectures and operating systems. We discussed test results based on 
custom-made benchmark data sets and also the Millennium II simulation, which is the
largest cosmological simulation currently available containing 10 billion particles.

Our future work will involve porting and running the parallelized Splotch on hybrid 
architecture computing systems containing several multiprocessor computers with CUDA 
enabled graphics boards, thus exploiting MPI and CUDA simultaneously. 
Several optimizations are also planned for our CUDA implementation, e.g.\ unrolling 
short loops for improved control flow or using local/shared memory to accelerate data fetching. 
Mechanisms for optimal load balance between CPU and the graphics processor and between several 
graphics processors when these are available should also be considered. We will 
investigate possibilities for designing advanced scheduling mechanisms to minimize 
the idle CPU times contained in the current implementation. Finally, we will explore the 
opportunities offered by the OpenCL library, in order to exploit a wider range of 
computing architectures.

\section*{Acknowledgments}
We would like to thank Mike Boylan-Kolchin for providing the Millenium II simulation data. 
Klaus Dolag acknowledges the support 
by the DFG Priority Programme 117. This work was also supported by the promising 
researcher's award, University of Portsmouth, and the HPC-EUROPA 2 
under the EC Coordination and Support Action Research Infrastructure Programme in FP7. 
Martin Reinecke is supported by the
German Aeronautics Center and Space Agency (DLR), under program 50-OP-0901, funded by the
Federal Ministry of Economics and Technology.

\section*{References}



\begin{thebibliography}{00}

\bibliographystyle{elsarticle-num}
\bibliography{master.bib}





\bibitem{sdss} http://www.sdss.org/

\bibitem{lofar} http://www.lofar.org/

\bibitem{2009MNRAS.398.1150B}http://www.mpa-garching.mpg.de/galform/millennium-II/index.html

\bibitem{lsst}http://www.lsst.org/lsst

\bibitem{iraf} http://iraf.noao.edu/

\bibitem{midas} http://www.eso.org/sci/data-processing/software/esomidas/

\bibitem{sao} http://tdc-www.harvard.edu/software/saoimage.html

\bibitem{gaia} http://astro.dur.ac.uk/~pdraper/gaia/gaia.htx/index.html

\bibitem{gnuplot} http://www.gnuplot.info/

\bibitem{supermongo} http://www.astro.princeton.edu/~rhl/sm/sm.html

\bibitem{idl} http://www.ittvis.com/


\bibitem{paraview} http://www.paraview.org/

\bibitem{aladin} http://aladin.u-strasbg.fr/

\bibitem{topcat} M. B. Taylor, TOPCAT and STIL: Starlink Table/VOTable Processing Software, ASP, ASPC 347 29T, 2005.

\bibitem{visivo1}M. Comparato, U. Becciani, A. Costa, B. Garilli, C. Gheller, B. Larsson and J. Taylor, 
PASP, Volume 119, 898-913, 2007.

\bibitem{visivo2}U. Becciani, A. Costa, V. Antonnuccio-Delogu, G. Caniglia, M. Comparato, 
C, Gheller, Z. Jin, M. Krokos, P. Massimino, PASP, 
Volume 122, 119-130, 2010


\bibitem{3dslicer} M. A. Borkin, N. A. Ridge, A. A. Goodman and M. Halle, astro-ph/0506604, 2005.

\bibitem{splash} http://users.monash.edu.au/~dprice/splash/index.html

\bibitem{visit} https://wci.llnl.gov/codes/visit/



\bibitem{2008NJPh...10l5006D}K.Dolag, M. Reinecke, C.Gheller, S. Imboden, 
New Journal of Physics, Volume 10, Issue 12, pp. 125006, 2008.

\bibitem{fraedrich2009}R. Fraedrich, J. Schneider, R. Westermann, 
IEEE Transactions on Visualization and Computer Graphics, vol. 15, no. 6, pp. 1251-1258, 
Nov./Dec. 2009, doi:10.1109/TVCG.2009.142

\bibitem{Szalay2008}T. Szalay, V. Springel and G. Lemson, http://arxiv.org/abs/0811.2055, 2008

\bibitem{mpi} http://www.mpi-forum.org/


\bibitem{cuda} http://www.Nvidia.com/object/cuda\_home.html

\bibitem{1985A&A...149..135M} J. Monaghan and J. Lattanzio, A\&A, 149, 135, 1985

\bibitem{1991par..book.....S} Physics of Astrophysics: Volume I Radiation, Shu F., 1991,
Published by University Science Books, 648 Broadway, Suite 902, New York, NY 10012


\bibitem{cudaprogguide} NVIDIA CUDA Programming Guide, Version 2.1 Beta, 10/23/2008, http://www.Nvidia.com

\bibitem{gadget} Springel V 2005 Monthly Notices of the Royal Astronomical Society 364, 1105-1134

\end{thebibliography}
\end{document}